\documentclass[letterpaper, 12pt, article]{ieeeconf}

\usepackage{times}
\usepackage[dvips]{color}
\usepackage{amssymb}
\usepackage{graphics,graphicx}
\usepackage[dvipsnames]{xcolor}
\usepackage{colortbl}
\usepackage{rotating}
\usepackage{bigstrut}
\usepackage{framed}
\usepackage{array}
\usepackage{cite}
\usepackage{float}
\usepackage{multirow,bigdelim}
\usepackage{multicol}
\usepackage{booktabs}
\usepackage{subfloat}
\usepackage{pgfplots}
\usepackage{subcaption}
\usepackage{microtype}
\usepackage{epstopdf}
\usepackage{wrapfig}
\usepackage{amsmath}
\usepackage{caption}
\usepackage{subfig}
\usepackage{cleveref}
\usepackage{balance}
\usepackage[switch]{lineno} 
\IEEEoverridecommandlockouts
\begin{document}
%\linenumbers
\newcommand{\squeezeup}{\vspace{-2.5mm}}
\newcommand{\argmax}[1]{\underset{#1}{\operatorname{arg}\,\operatorname{max}}\;}
\newcommand{\mytilde}{\raise.17ex\hbox{$\scriptstyle\mathtt{\sim}$}}
\newcommand*{\TitleFont}{%
	\usefont{\encodingdefault}{\rmdefault}{b}{n}%
	\fontsize{16}{19.2}%
	\selectfont}
%\includepdf[pages=1-,pagecommand={\thispagestyle{plain}}]{<pdffile>}
% paper title
%\newcommand{\specificthanks}[1]{\@fnsymbol{#1}}% Inserts a specific \thanks symbol

%\title{\LARGE \bf
\title{\TitleFont
Advantages of EEG phase patterns for the detection of gait intention in healthy and stroke subjects \\
}

\author{Andreea~Ioana Sburlea\textsuperscript{1,2,*} \quad Luis Montesano\textsuperscript{1,2} \quad Javier Minguez\textsuperscript{1,2}
\thanks{
\textsuperscript{1} University of Zaragoza (DIIS), Instituto de investigaci\'on en ingenier\'ia de Arag\'on (I3A);
\textsuperscript{2} Bit\&Brain Technologies S.L., Zaragoza, Spain. 
\newline
\textsuperscript{*} Andreea Ioana Sburlea, Bit\&Brain Technologies S.L., Paseo Sagasta 19, 50001, Zaragoza, Spain. \newline
% Corresponding author's 
E-mail:~andreea.ioana.sburlea@gmail.com
}}

\maketitle

\begin{abstract}
One use of EEG-based brain-computer interfaces (BCIs) in rehabilitation is the detection of movement intention.
In this paper we investigate for the first time the instantaneous phase of movement related cortical potential (MRCP) and its application to the detection of gait intention.
We demonstrate the utility of MRCP phase in two independent datasets, in which 10 healthy subjects and 9 chronic stroke patients executed a self-initiated gait task in three sessions. 
Phase features were compared to more conventional amplitude and power features.
The neurophysiology analysis showed that phase features have higher signal-to-noise ratio than the other features.
Also, BCI detectors of gait intention based on phase, amplitude, and their combination were evaluated under three conditions: session specific calibration, intersession transfer, and intersubject transfer.
Results show that the phase based detector is the most accurate for session specific calibration (movement intention was correctly detected in 66.5\% of trials in healthy subjects, and in 63.3\% in stroke patients). However, in intersession and intersubject transfer, the detector that combines amplitude and phase features is the most accurate one and the only that retains its accuracy (62.5\% in healthy subjects and 59\% in stroke patients) w.r.t. session specific calibration. Thus, MRCP phase features improve the detection of gait intention and could be used in practice to remove time-consuming BCI recalibration.

%
%The neurophysiology analysis showed that phase features have higher signal-to-noise ratio than the other features.
%In BCI detection
%We have found that the phase based detector is the most accurate in intrasession evaluation. 
%However, in intersession and intersubject transfer, the detector that combines amplitude and phase features is the most accurate one and the only that retains its accuracy. 
%Using the phase based BCI in intrasession condition, movement intention was correctly detected in 66.5\% of trials in healthy subjects, and in 63.3\% in stroke patients. 
%For the combination of features we attained 61\% in intersession transfer in healthy subjects and 58.3\% in stroke patients. The intersubject transfer shows similar performance, 62.5\% in healthy and 59\% in stroke patients. 
%%Furthermore the detector based on a combination of features detect earlier de movement intention than the other detectors.  
%These results show that MRCP phase features improve the detection of gait intention and could be used in practice to remove time-consuming BCI recalibration.

\end{abstract}

\begin{keywords}
EEG, BCI, gait intention, MRCP, intersession transfer, intersubject transfer, instantaneous phase

\end{keywords}

\IEEEpeerreviewmaketitle
%--------------------------
\section{Introduction}
 
Brain computer interfaces (BCIs) as a rehabilitaton tool have been used to restore functions in patients with gait impairments by actively involving the central nervous system to trigger prosthetic devices according to the detected intention to walk~\cite{belda2011rehabilitation}. 
One of the most investigated neural correlates of movement intention, as imaged by electroencephalogram (EEG), is the movement related cortical potential (MRCP)~\cite{hh1965changes, shibasaki2006bereitschaftspotential}. 
The relation between MRCP and movement intention has been extensively studied with EEG-based BCIs in the context of self-paced lower limb movements and gait~\cite{wheaton2007does, presacco2011towards, gwin2010removal, gwin2011electrocortical,jiang2014accurate, jiang2014brain, bulea2014sitting}.

All studies that used MRCP information for the detection of movement intention explored the amplitude representation of the neural correlate~\cite{niazi2011detection, lew2012detection, do2006movement, niazi2013detection, garipelli2013single, jiang2014brain, sburlea2015continuous, sburlea2015detecting}.
In particular, the amplitude of the MRCP has been used to detect gait intention in healthy subjects within session~\cite{jiang2014brain} and between sessions ~\cite{sburlea2015continuous}. 
In the frequency domain, the neural correlate of movement intention is the event related (de)-synchronization (ERD/S) in mu and beta bands~\cite{pfurtscheller1979evaluation, pfurtscheller2006mu, formaggio2013modulation, nam2011movement, severens2012feasibility, wagner2014s}. 
An improvement in performance within session and between sessions was obtained by combining the two motor intention neural correlates~\cite{ibanez2014detection, savic2014movement, sburlea2015continuous, sburlea2015detecting}.  
Recent work has been carried out to eliminate or to diminish the time required for BCI calibration in intersession and intersubject transfer learning~\cite{tu2012subject, lotte2010learning, krauledat2008towards, arvaneh2012omitting, arvaneh2013eeg, fazli2009, fazli2009a,fazli2011l, garipelli2013single, lottesignal, fazli2015learning, sburlea2015continuous, sburlea2015detecting, sburlea2015adaptation, lopez2015brain}.

Although phase patterns have been investigated before  in the context of event related-potentials in theta and alpha band, the MRCP phase representation ($0.1-1$~Hz delta oscillations) has not been explored yet.  
Oscillatory activity in theta and alpha bands revealed that instantaneous phase patterns can contain information about the response to external stimuli (visual or auditory event-related potentials)~\cite{busch2009phase, ng2013eeg}. Furthermore, they found that the information that can be discriminated by firing rates can also be discriminated by phase patterns, but not by power.
In the delta frequency band, in which the MRCP is also observed, recent studies~\cite{saleh2010fast, stefanics2010phase} have reported that the phase patterns are linked to the anticipation of visual and auditory stimuli. However, it remains unknown whether the phase patterns of the slow oscillations in the delta band are as discriminable as the  amplitude patterns, as shown before for faster oscillations~\cite{busch2009phase, ng2013eeg}. 
%ory mechanisms, showing a larger synchronization in the p and can predict behavioral reaction s carry information about the reaction time to auditory or visual stimuli.
% can anticipate the target stimuli, but not the non-target ones. 
%
In studies about movement related neural correlates, phase patterns have been investigated mainly in the mu band as a metric of connectivity between brain areas~\cite{daly2012brain, wei2007amplitude, he2012combination, hamner2011phase,wagner2013robot, hsu2014enhancing}. 

In this paper we present first, the decomposition of MRCP amplitude into instantaneous phase and power, and second, we present three detectors of gait intention based on (a) MRCP amplitude, (b) MRCP instantaneous phase and (c) a combination of MRCP amplitude and MRCP instantaneous phase.
% the outputs of the previous two detectors.
We performed our analysis on two independent datasets recorded previously in separate studies, one in healthy subjects~\cite{sburlea2015continuous} and another in chronic stroke patients~\cite{sburlea2015detecting}. The group of healthy subjects is not a control group for the patients.   
Both groups of subjects underwent three self-paced walking sessions with one week between them. 
We show the applicability of the detectors of walking intention in three cases: within session, between sessions and between subjects, without recalibration. 

Our findings indicate that MRCP instantaneous phase shows higher signal-to-noise ratio than amplitude and instantaneous power. This characteristic of phase yields higher detection accuracy of pre-movement state within session, as well as an earlier detection of movement intention compared to the MRCP amplitude based detector.
Previous work~\cite{xu2014enhanced} has shown that the latency of the detection of movement intention has an important effect on neuroplasticity and relearning. Small detection latencies relative to the movement onset aid functional recovery in rehabilitation therapies. 
Furthermore, patients are fatigue-prone in prolonged and repetitive therapy sessions. Thus, it would be beneficial to remove the need for session- and subject-specific BCI recalibration. 
By combining MRCP amplitude and phase information, without recalibration, we attained a detector with a more robust performance between sessions and subjects, outperforming the detectors that use only one type of information. 
%

%	
%-------------------------X
\section{Materials and methods}
%----------

\subsection{Description of the datasets}
%\subsection {Participants and experimental procedure}
Two previously recorded datasets  from two separate studies~\cite{sburlea2015continuous, sburlea2015detecting} have been used in the current analysis.
More information about the demographic and clinical information of the participants, the experimental procedure and the data acquisition can be found in the Supplementary Materials.   

\subsubsection{Dataset~1} 
Ten volunteers (six males and four females, mean age~=~26.4~years, \textit{SD}~=~4.8~years) participated in the experiment described in~\cite{sburlea2015continuous}. 
%They were recruited from Bit\&Brain Technologies in Zaragoza, Spain. 
All subjects were healthy without any known neurological anomalies and musculo-skeletal disorders.
%These subjects were part of the dataset described in~\cite{sburlea2015continuous}.

\subsubsection{Dataset~2}
Nine chronic stroke patients (six males and three females, mean age~=~59.7~years, \textit{SD}~=~11.3~years) participated in the experiment presented in~\cite{sburlea2015detecting}. 
%They were recruited from the Clinic of University Rey Juan Carlos in Madrid, Spain. 
Demographic and clinical information of the participants can be found in~\cite{sburlea2015detecting}.

\begin{figure}[h!]
   \centering
    \includegraphics[width=0.45\textwidth]{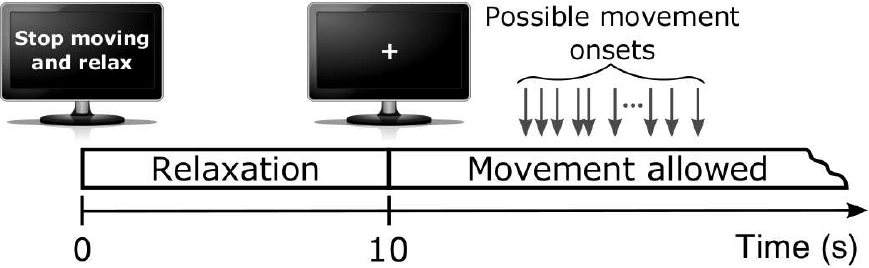}%{Figure0_Experiment_protocol.eps} % Figure1_Experiment_protocol
 	\caption{Structure of one trial of the experimental protocol. }
\label{fig:Protocol}
\end{figure}

The experimental protocol in the two datasets was approved by the ethical committee of the HYPER project (approval number 12/104)~\footnote{This project is part of the Spanish Ministry of Science Consolider Ingenio program, HYPER (Hybrid Neuroprosthetic and Neurorobotic Devices for Functional Compensation and Rehabilitation of Motor Disorders) - CSD2009-00067. } and all subjects gave written informed consent before participating in the experiments. 
Figure~\ref{fig:Protocol} illustrates the structure of one trial of the experimental protocol. 

In each of the datasets, the experimental protocol was performed in three sessions with one week between them. Each session had a total of 100 trials. In each trial, subjects were instructed with visual cues (in the healthy subjects) or auditory cues (in the stroke patients) to relax for the first $10$~s, then start walking whenever they wanted, but not before waiting at least $1.5$~s after the instruction cue. After each ten trials there were break intervals with a duration adjusted to the need of the participants.
The experimental protocol was slightly modified in the second dataset, for the stroke group, to accommodate flexible pauses for patients to regain balance. 

%The protocol was similar to the one shown in Chapter 2 with the following differences: the visual cues were replaced with auditory cues, the footswitch was replaced with EMG sensors for the detection of muscular activation and longer intervals were introduced for the initiated of motion. The EMG sensors were located on the tibialis anterior muscles of both legs. These changes were incorporated due to the patients' condition (difficulty in maintaining a fixed gaze during walking and the presence of uncontrolled movement, such as tremor). More information about the structure of the protocol can be found in Chapter 2.

%For a more detailed description of the protocol, see~\cite{sburlea2015continuous} for dataset~1, and~\cite{sburlea2015detecting} for dataset~2. 

%The experimental protocol was the same during all three sessions.  
%\begin{figure}[h]
%   \centering
%%	\vspace{-2.45cm}
%    \includegraphics[width=0.45\textwidth]{Figure0_Experiment_protocol}%{Protocol_JNE1_onesession}%{fig_1}
% 	\caption{Protocol of the experiment. }
%\label{fig:Protocol}
%\end{figure}
%%\vspace{-2.45cm}

%-------------------------X
%\subsection{Data acquisition}
In the two datasets EEG data was recorded using 
%a TMSi REFA amplifier and 
30 EEG water based electrodes (from TMSi, Enschede, The Netherlands) located according to the 
%. The EEG sensors were located at Fp1, Fpz, Fp2, F7, F3, Fz, F4, F8, FC5, FC1, FC2, FC6, T7, C3, Cz, C4, T8, CP5, CP1, CP2, CP6, P7, P3, Pz, P4, P8, POz, O1, Oz, O2, following the 
10/10 international system~\cite{jurcak200710}. The ground was placed on the right wrist and two sensors located on the ear lobes have been used for average linked ears reference.  Electromyographic (EMG) activity and footswitch presses were recorded during the experiment with healthy subjects and only the EMG activity, during the recordings with patients. 
%For details about the data acquisition process see~\cite{sburlea2015 continuous, sburlea2015detecting}.
%-------------------------X
\subsection{Data processing}
EEG data was filtered with a Butterworth second order zero-phase shift band-pass filter at $0.1-30$~Hz. 
EMG data was filtered at $100-125$~Hz using a Butterworth second order zero-phase shift band-pass filter.
For the data in the first dataset, the onset of motion was computed as being half a second before the footswitch release.  
For the second dataset, EMG data was used for the computation of movement onset, since the acquisition of the footswitch activity was impractical in the experiment with stroke patients due to different levels of mobility impairments.
The EMG data was Hilbert transformed and a threshold of 10\% of the highest value from the averaged EMG power across trials was computed. The onset of motion was defined as $100$~ms before the threshold crossing.
%For a detailed description of the onset detection technique in dataset~2, see~\cite{sburlea2015detecting}.
The trials in which the onset of motion was detected before the preparation of movement cue were rejected as considered artifactual due to erroneous execution of the experimental protocol.
No trials have been rejected in the first dataset.
After trial rejection, in the second dataset, there were in average across subjects and sessions 96 remaining trials, in a range between 86 and 100 trials. 
  
In both datasets, the remaining trials were segmented into $6$~s long epochs according to the movement onset (each epoch lasted from $-6$ to $0$~s relative to the movement onset).
EEG data in dataset~1 was processed with FastICA~\cite{bingham2000fast} for artifact correction, as described in~\cite{sburlea2015continuous}. For the EEG data in dataset~2, trials that 
were contaminated by large, infrequent artifacts were removed using the joint probability function from EEGLAB 13.3.2 toolbox \cite{delorme2004eeglab}. Next the remaining trials were processed with FastICA. 
%A more detailed description of the procedure can be found in~\cite{sburlea2015detecting}. 

Movement related cortical potentials (MRCP) were analyzed as EEG neural correlates of gait intention.
For the analysis, EEG data was band-pass filtered with a zero-phase shift Butterworth second order filter at $0.1-1$~Hz. 
As shown in \cite{garipelli2011single}, this low-frequency filter has reliable characteristics for the analysis of MRCPs. 

% our proposed methodology 
The MRCP amplitude signals $s(t)$ were decomposed  into instantaneous amplitude and instantaneous phase using the analytic representation~\footnote{Note that the MRCP amplitude signals are the original signals before Hilbert transform, while the instantaneous amplitude is a component of the MRCP signals obtained after Hilbert transform.} 
\begin{equation}
z(t) = s(t) +  j\mathrm{HT}(s(t))
\end{equation}

where HT($s(t)$) is the Hilbert transform of the signal $s(t)$, defined as
\begin{equation}
\mathrm{HT}(s(t)) = \frac{1}{\pi} \mathrm{P.V.}\int_{-\infty}^{\infty} \frac{s(t)}{t-\tau}\mathrm{d}\tau
\label{eq:Hilbert}
\end{equation}

where P.V. denotes Cauchy principal value.
Therefore, the analytic signals $z(t) = A(t)e^{j\phi(t)}$ were obtained for each electrode.
The instantaneous amplitude is expressed by $A(t)$ and the instantaneous phase by $\phi(t)$.
 A convenient mathematical representation of phase angles as vectors with unit length is given by Euler's formula ($e^{j\phi}~=~cos(\phi)~+~jsin(\phi)$). 
To characterize the inter-trial phase consistency between narrow-band signals, phase locking value (PLV) is frequently used,
\begin{equation}
PLV = \frac{1}{N}|\Sigma_1^N e^{j\phi}|
\end{equation}

where \textit{N} represents the total number of trials~\cite{lachaux1999measuring}. 
This measurement is performed by averaging the vectors (not their phase angles) and taking the length of the average vector as a measure of uniformity in the polar space.
PLV ranges between zero and one, where zero indicates completely uniformly distributed phase angles and one indicates completely identical phase angles \cite{cohen2014analyzing}.
% Terminology definition

\label{classify}
\subsection{Neurophysiological analysis}
We conducted the neurophysiological analysis at two levels, using the first session of the two datasets. First, at a scalp level, we performed a temporal analysis to investigate three types of features:
\begin{enumerate}
\item MRCP amplitude features of the signal filtered in $0.1-1$~Hz frequency band; 
\item MRCP phase features: instantaneous phase features obtained with Hilbert transform, and PLV as a summary statistic for single-trial instantaneous phase features; 
\item MRCP (instantaneous) power features obtained with Hilbert transform as the logarithm of squared instantaneous amplitude values and multiplied by 10 to yield the decibels scale.
\end{enumerate}

Second, for channel Cz, which exhibits relevant patterns related to gait intention over the motor areas~\cite{jiang2014brain, sburlea2015continuous}, we conducted a statistical analysis to assess the signal-to-noise ratio of the features. 
Therefore, we computed the effect size for each type of feature relative to the baseline interval (between $-5.5$ and $-4$~s relative to the movement onset).
More precisely, we computed the average of the three features over trials, for each subject. Next we calculated the baseline mean and standard deviation. Finally we quantified the effect size, for each feature type, by subtracting from the grand average activity (over subjects) the grand average baseline, and dividing by the standard deviation of the baseline.  

We performed for each of the datasets, a statistical analysis at neurophysiological level, using pairwise Wilcoxon signed rank tests. For the analysis we used the absolute values
%\footnote{The effect size for amplitude representation is negative due to the decrease in magnitude of the feature. However a direct comparison of the effect sizes (not of their absolute values) would be meaningless.} 
of the effect sizes of the three features (amplitude, instantaneous phase and power) in channel Cz. We used a non-parametric test to avoid inappropriate assumptions about the distribution of the features. We chose to compare the absolute values of the effect size due to the differences in sign between them. Next we corrected for multiple comparisons using Bonferroni-Holm correction~\cite{holm1979simple}.
We evaluated the statistical significance between the effect sizes of the three representations in the intention of motion interval ($-1.5, 0$s).  

\subsection{Feature extraction and classification} 
At the classification level, we studied the MRCP amplitude and instantaneous phase features. The instantaneous phase features were decomposed into sine and cosine features according to Euler's formula.
Features for both processes were extracted from ten electrodes with a one second long sliding window in steps of $125$~ms. The ten electrodes (F3, Fz, F4, FC1, FC2, C3, Cz, C4, CP1 and CP2) were located over the precentral, central and postcentral motor and sensorimotor cortex.  
Windows between $-6$ and $-1.5$~s were labeled as relaxation state and those from $-1.5$ to $0$~s as pre-movement state. 
We analyzed a total of 41 windows per trial out of which the last 8 belonged to the pre-movement class.  Before training the classifier,  features were normalized to unit Euclidean length.

For the classification we used a support-vector machine (SVM) with radial basis function (RBF) kernel.
This method has two hyperparameters: the RBF parameter $\gamma$ and the regularization parameter \textit{C}. We used this method due to the nonlinear nature of the instantaneous phase features.
%Gamma represents the influence of a single training sample on the selection of support vectors, with high values meaning closer to the chosen support vectors. Cost is a tradeoff misclassification of the training samples against the simplicity of the surface. A higher c would tend to correctly classify each sample and select more support vectors.  
  
Three detection models were built based on different sets of features: 
(1)~Amplitude model, a single-view detection model based on low-frequency MRCP amplitude features,
(2)~Phase model, a single-view detection model based on MRCP instantaneous phase features, and (3) Amplitude~+~Phase model, a multi-view detection model. The first layer of the multi-view model contains, as features, the outputs of the two single-view detectors, Amplitude model and Phase model, while in the second layer we used linear discriminant analysis to combine the two single-view models.

Model selection and artifact cleaning procedures were performed on each fold inside a $5\times5$-fold nested chronological cross-validation~\cite{lemm2011introduction}. During model selection, the SVM hyperparameters, RBF parameter $\gamma$ and regularization parameter \textit{C}, were selected using a $5\times5$ grid search in the inner loop of the cross-validation. The search range for both parameters was between $2^{-5}$ and $2^{5}$. 
The probability threshold was automatically selected in the inner loop of the cross-validation as the one that maximizes the performance on the validation set of the inner loop.    	
In the outer loop we computed the classification performance.

%--------
\subsection{Evaluation}
\label{metrics}
%Model selection and artifact cleaning procedures were performed on each fold inside a 5x5-fold nested chronological cross-validation. During model selection, the SVM hyperparameters, RBF parameter $\gamma$ and regularization parameter C, were selected in the inner loop of the cross-validation. In the outer loop we computed the classification performance.
%The probability threshold was automatically selected in the inner loop of the cross-validation as the one that maximizes performance on the inner loop validation set.    	

We assessed the performance of the detection models in intra- and intersession conditions.
In the intrasession condition, we evaluated the detection models in a $5\times5$-fold chronological cross-validation using subject-specific data within a session.   
%
%, averaged across trials and sessions, for subjects of both groups.    
%
In the intersession transfer condition we trained the detection models with data from one session and tested on all the data of a subsequent session, respecting the chronological order of the sessions in each of the datasets.

For the assessment of the three detection models (Amplitude model, Phase model and Amplitude~+~Phase model) in the intra- and intersession conditions, we used two metrics of performance: at the level of window and at the level of trial (sequence of sliding windows).
 
At the window level we measured the performance of correctly classified windows using receiver operating characteristic (ROC) curves. We conducted this analysis for all the models in intrasession and intersession conditions. 
In order to asses the performance while taking into account the empirical chance level, we have also used the Cohen's kappa~\cite{billinger2013significant}.
%We computed the chance level at the level of windows using Cohen's kappa~\cite{billinger2013significant}.
% and we attained, in healthy subjects the average chance level of 59$\pm$2\%, $\kappa$~=~0.68$\pm$0.1, and in stroke patients, the average chance level of 58$\pm$3\%, $\kappa$~=~0.63$\pm$0.12. 

At the trial level we assessed the performance of the models, in all the conditions as the percentage of correctly classified trials~\cite{sburlea2015continuous}. This metric is dependent on the labeling sequence of the sliding windows. Pre-movement class ($-1.5$~s to the onset of movement, equivalent to the last 8 sliding windows) was defined as the positive class. Thus a correct trial is defined according to the following logical expression:
\begin{equation}
\mathrm{correct~trial} = \bigg(\bigwedge_{i=1}^{33} \neg{\mathrm{FP}}_i\bigg) \land \bigg(\bigvee_{i=34}^{41} \mathrm{TP}_i\bigg) 
\end{equation}

where FP and TP are the false and the true positive windows, respectively, and \textit{i} represents the index of the window. The symbol $\bigwedge$ stands for the logical product, while the symbol $\bigvee$ stands for the logical sum of the windows. In other words, the logical expression defines as correct a trial in which there are no false positive windows and there is at least one true positive window.
This metric gives a conservative estimate of performance compared to the accuracy computed at the level of windows. Furthermore, this metric is more meaningful in rehabilitation scenarios, in which neural plasticity and implicitly functional recovery are determined by the correctly detected self-initiated (attempts of) movement of the participants~\cite{belda2011rehabilitation}. 

We conducted statistical tests of the trial-based performance using repeated measures two-way ANOVA, with factors ``detection model type'' (amplitude, phase, combined amplitude and phase) and ``evaluation type'' (intra- and intersession), in which the averaged performance across sessions for individual subjects are the repeated measures. We tested for the sphericity of the data using the Mauchly's test. We evaluated the difference between intra- and intersession conditions using pairwise $t$-tests. We corrected for multiple comparisons using the Bonferroni-Holm correction. 

%!!! Chance level window based and Cohen's kappa
%Probably the following paragraph ca be removed or rephrased

%The performance obtained as percentage of correctly classified trials were above chance level, which was computed by randomly interchanging the labels of a whole sequence of sliding windows (33 for the rest class and 8 for the pre-movement class) between classes. Due to class imbalance and to a long sequence length (41 windows), the chance level was 4\%. 

% More specifically if it contains at least one true positive window (i.e. pre-movement state detection) in the [$-1.5$; 0~s] time interval and has no false positive windows in the [$-6$; $-1.5$s] time interval.

%In the intrasession (intrasubject) condition, we evaluated the models in a $5\times5$-fold chronological cross-validation using subject-specific data within a session.   
%%
%%, averaged across trials and sessions, for subjects of both groups.    
%%
%In the intersession transfer condition we trained the models with data from one session and tested on all the data of a subsequent session, respecting the chronological order of the sessions in each of the datasets.

In addition to the transfer between sessions we performed the intersubject evaluation in a leave-one subject out cross-validation within each session using the trial-based metric. In each dataset, we trained the models with the data of all but one subject and evaluated on the data of the left out subject. We repeated the procedure, evaluating for each subject of the two groups. The intrasubject evaluation was equivalent with the intrasession evaluation. 
%For the evaluation of the transfer conditions we used all three detection models presented above. 

We evaluated the difference between intra- and intersubject conditions using a pairwise Wilcoxon signed rank test. We corrected for multiple comparisons using the Bonferroni-Holm correction. 

Furthermore, we performed the analysis of the detection time (latency) for the intrasession condition based on the probability outputs of the correctly classified trials, for all the detection models and subjects. 
The detection time was computed as the time point of the first true positive window (first window correctly classified as movement intention). 
We performed a pairwise Wilcoxon signed rank statistical test of the detection time between the three detection models, for each dataset and corrected for multiple comparisons using Bonferroni-Holm correction.

The statistical analysis was conducted in R~\cite{ihaka1996r}, while the rest of the analysis was performed in Matlab R2011a (MathWorks, Massachusetts, US).
% statistics for the latency analysis (Wilcoxon signed rank test)
% statistics for the neurophysiology analysis (nonparametric Wilcoxon test). 

%----------------------
\section{Results}
\label{sec:results}

\subsection{Neurophysiological analysis of healthy subjects}
\label{subsec:NeuroResults}
%\subsubsection{Healthy subjects}
%========= HEALTHY ====================

\begin{figure*}
\centering
\includegraphics[width=1\textwidth]{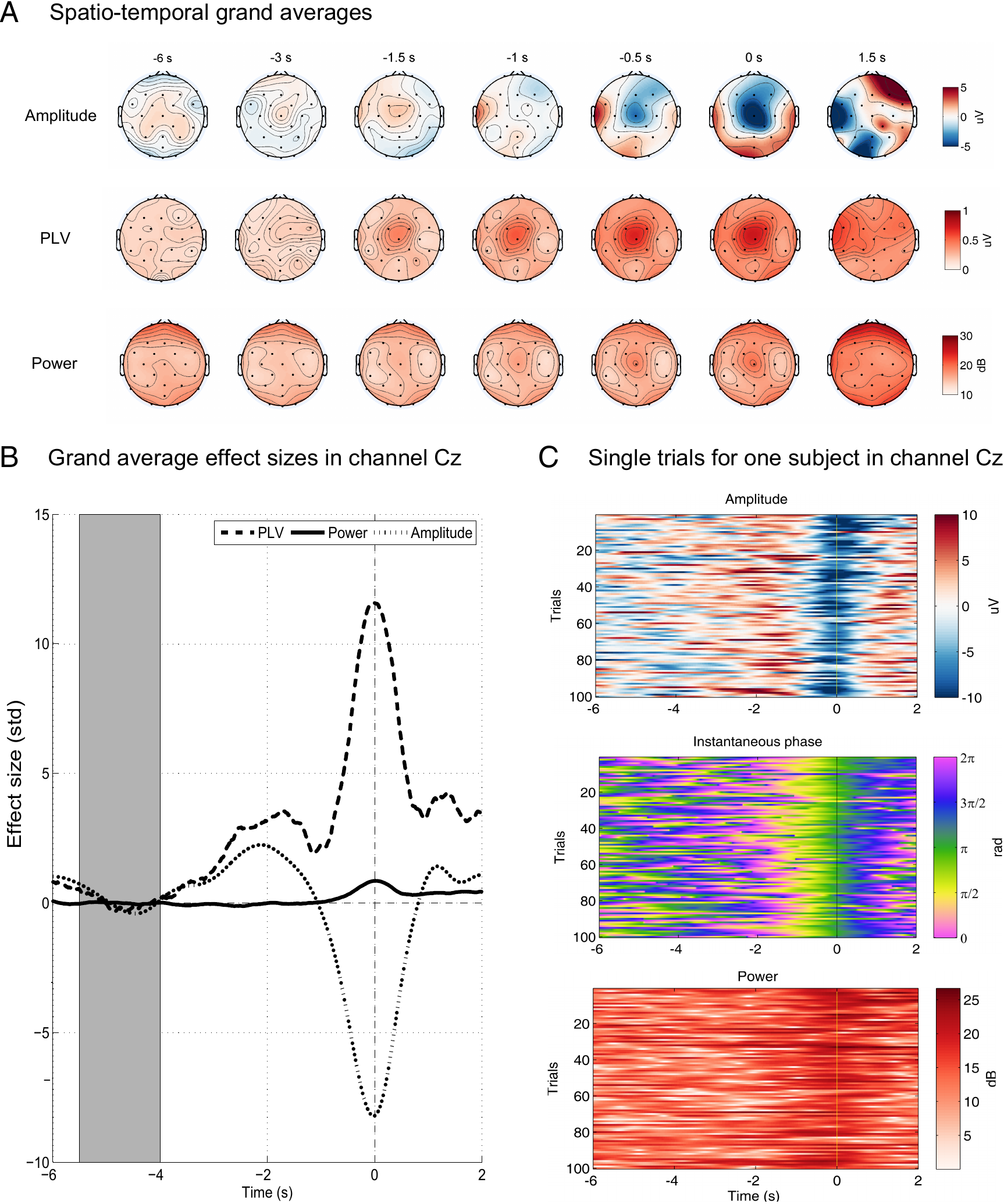}%{rev_healthy_neuro.eps}%{Figure2_Neuro_healthy.eps} % Figure2_Neuro_healthy
\caption{Healthy subjects neurophysiological results. 
A. Grand average EEG correlates of gait intention in a time interval between $-6$ and $1.5$~s relative to the motion onset. 
Top panel: Movement related cortical potential (MRCP) amplitude representation; 
Middle panel: Phase-locking value (PLV) over task repetitions; 
Bottom panel: Power on decibels scale. 
B. Effect sizes for each of the three features quantified in standard deviation units in channel Cz for the grand average across subjects. The gray area indicates the baseline used for the computation of the effect size.
C. Single trials illustration of the features in central channel Cz for one subject across multiple repetitions. 
The vertical dashed line indicates the onset of motion.}
\label{fig: Figure1_Neuro_Healthy}
\end{figure*}

Figure~\ref{fig: Figure1_Neuro_Healthy} shows the low-frequency EEG correlates of gait intention in healthy subjects (dataset 1).
Figure~\ref{fig: Figure1_Neuro_Healthy}A presents the MRCP features in grand average over subjects. On the first panel the MRCP amplitude pattern is observed as a negative deflection in the activity recorded by electrodes located over the motor cortex, starting $1$~s prior to the movement onset. 
The middle panel shows the PLV or the phase synchronization.
% computed across repetitions of the task (trials) and averaged over subjects.
An increase in PLV over the motor cortex electrodes indicates a higher synchronization in the phases patterns starting $1.5$~s prior to the movement onset.
The bottom panel presents the evolution of low-frequency power across time. In the proximity of movement onset there is a slight increase in power. 

Figure~\ref{fig: Figure1_Neuro_Healthy}B presents the size of the effect of the movement intention on the three features across channel Cz. This channel is chosen because it presents a relevant pattern over motor areas for all types of features~\cite{jiang2014brain, sburlea2015continuous}.  The estimation of the effect size was done relative to the baseline interval ($-5.5,-4$)~s with respect to the movement onset. 
In the intention of motion interval ($-1.5,0$)~s, PLV has the largest average effect size (5.25 standard deviations), while the amplitude of the MRCP has a lower average effect size (3.23 standard deviations). Power shows the smallest effect size (0.26 standard deviations).

We evaluated the statistical significance between the absolute value of the averaged effect size of the three representations (amplitude, instantaneous phase and power) in the intention of motion interval ($-1.5,0$)~s using a pairwise Wilcoxon test.
After the Bonferroni-Holm correction we found a statistically significant difference between instantaneous phase and power ($p$~=~0.002), as well as between amplitude and power ($p$~=~0.027). However, the difference between instantaneous phase and amplitude was not statistically significant ($p$~=~0.556).

%At the movement onset, PLV has the largest effect size, 11.6 standard deviations, while the amplitude of the MRCP has a lower effect size of 8.2 standard deviations. Power shows the smallest effect size of only 0.8 standard deviations.

Figure~\ref{fig: Figure1_Neuro_Healthy}C illustrates the behavior of the three types of features in single trials over the central channel Cz for one representative subject.
The course of the MRCP amplitude feature across trials is shown in the top panel.
A decrease in amplitude is visible $1$~s before the onset of motion. 
The middle panel presents a synchronization of instantaneous phase across trials.
After $1.5$~s prior to the movement onset, there is an ascending trend in phase from $\pi$/2 to $\pi$. 
Next, in the bottom panel power shows a less robust pattern across trials compared to the MRCP instantaneous phase or amplitude representations.

\subsection{Evaluation of detection models in healthy subjects}
\subsubsection{Intra- and intersession evaluation}
We evaluated the detection performance in healthy subjects during the intra- and intersession conditions using two metrics: window-based and trial-based. 
%Figure~\ref{fig:detection_models10} presents the detection performance attained in healthy subjects using the two metrics: window-based (in panel A) and trial-based (in panel B). Next, in panel C, we report the trial-based performance for intra- and intersubject evaluation. Finally, panel D shows the evaluation of detection time for the three detection models in the intrasession evaluation.   

\begin{figure*} % Healthy
\begin{tabular}{cc} % % FIGURE 4 a b
  \includegraphics[scale = 1]{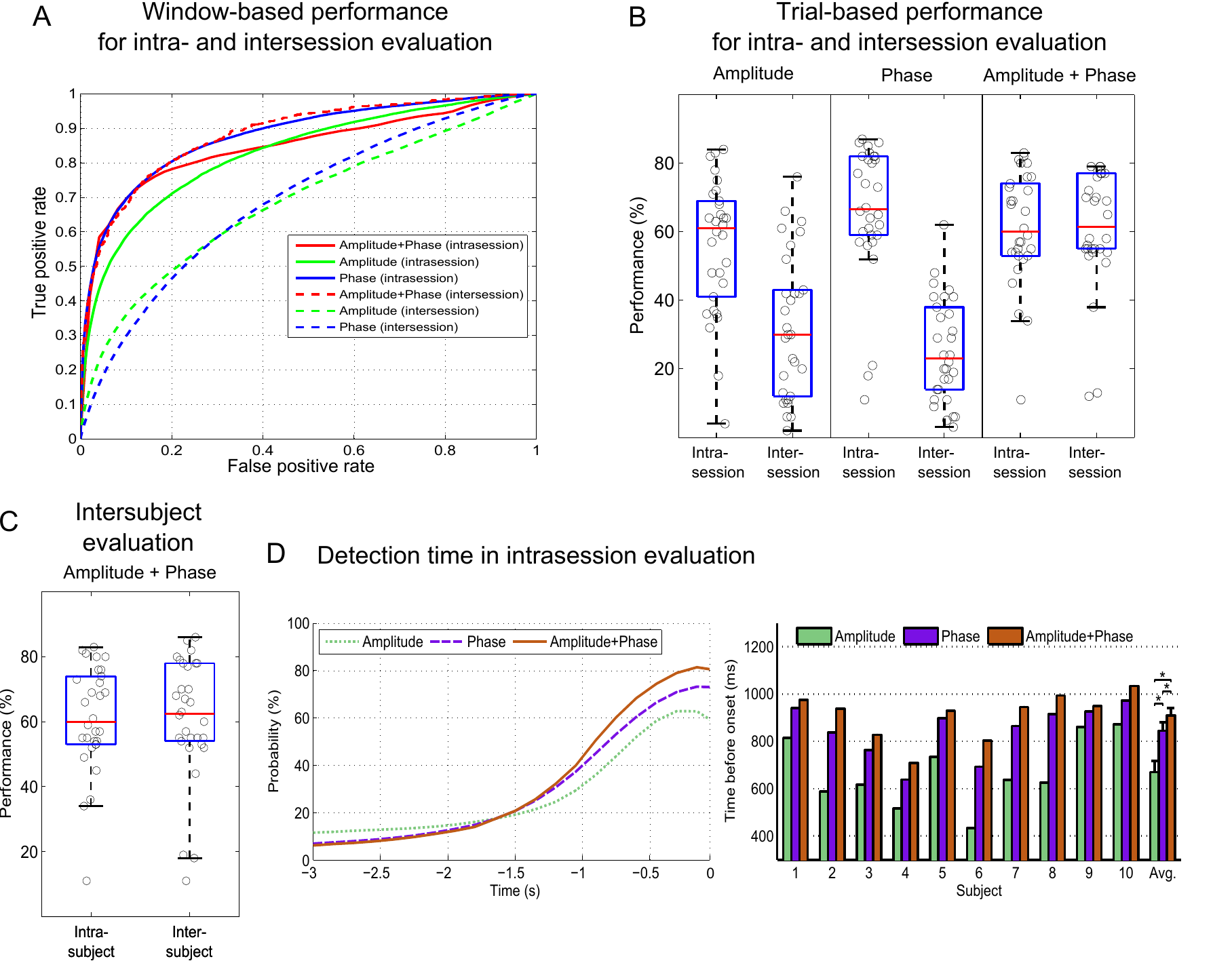}%{rev_Figure_det_healthy.eps}%{rev_Fig_perf_healthy.eps}% \qquad   %\includegraphics[width=0.46\textwidth]{Figure4b_Decoding_strokes.eps} \\ 
\end{tabular}
\caption{Performance detection of gait intention in healthy subjects using three models: Amplitude, Phase and Amplitude~+~Phase, in intrasession and intersession evaluations, computed in intra- and intersession with two metrics: 
A. window-based metric and B. trial-based metric. The circles mark subject and session specific performance. C. Trial-based intra- and intersubject evaluation. D. Detection time evaluation of the three detection models in intrasession condition. 
%
%Performance as percentage of correctly classified trials. A. Healthy subjects group. B. Stroke patients group. Left representation from both A and B panels shows the scores attained with the MRCP amplitude model during intrasession and intersession conditions. The middle representation in both panels show the MRCP phase model scores in intrasession and intersession conditions.  The right representation in both panels shows the scores of the multi-view model that combines the MRCP amplitude and MRCP phase single-view models, in intrasession and intersession conditions. The circles mark subject specific performance.
}
\label{fig:detection_models10}
\end{figure*}

\paragraph*{Window-based performance}
Figure~\ref{fig:detection_models10}A presents the averaged ROC curves  across subjects for the three detection models, in two evaluation conditions, intrasession and intersession. 
To quantify the window-based performance in our unbalanced experimental design, we computed the area under the curve (AUC) as a summary statistic of the presented ROC curves.
During intrasession the highest averaged AUC was attained for the Phase model ($M$~=~0.88, $SD$~=~0.09), while the lowest AUC was obtained using the Amplitude model ($M$~=~0.82, $SD$~=~0.09).
For the intersession, the highest AUC was obtained using the multi-view model ($M$~=~0.87, $SD$~=~0.09), and the lowest AUC was attained using the Amplitude model ($M$~=~0.69, $SD$~=~0.16).
We attained the average empirical chance level of 59$\pm$2\%, with Cohen's $\kappa$~=~0.68$\pm$0.1.
From these findings we conclude that in the group of healthy subjects, the Phase model is the most reliable in intrasession, while in intersession the Amplitude~+~Phase model attains higher performance relative to the other detection models.  

\paragraph*{Trial-based performance}
Figure~\ref{fig:detection_models10}B shows the performance of the three detection models at trial level, as percentage of correctly classified trials, in intrasession and intersession conditions.
% We present graphically in the figure the distribution in performance, by marking with circles the subject- and session-specific performance. 

In intrasession evaluation the Phase model attains the highest percentage of correctly classified trials ($Mdn$~=~66.5\%, range~=~11\% - 87\%), followed by the Amplitude model ($Mdn$~=~61\%, range~=~4\% - 84\%) and by the Amplitude~+~Phase model ($Mdn$~=~60\%, range~=~11\% - 83\%). 

%During intrasession, the Phase model has the highest median score 66.5\% in a range between 11\% and 87\% compared to the results of other detection models. 
%%the model based on amplitude features 61\% in a range between 4\% and 84\%.
%The intersession transfer performance attained with the Amplitude model and with the Phase model shows a large decrease in performance. 
%The median score obtained using the Phase model for intersession transfer is 23\% in a range between 3\% and 62\%. 
%%
%The median score for the Amplitude model in intersession transfer is 30\% in a range between 2\% and 76\%. 
% 
%Moreover, during intrasession, the model that combines the amplitude and phase information yields the highest percentage of correctly classified trials (Mdn = 60\%, range = 11\% - 83\%). 
During the intersession transfer condition, the multi-view model attained similar performance ($Mdn$~=~61\%, range~=~12\% - 79\%) to the intrasession evaluation. 
The intersession transfer performance attained with the Amplitude model and with the Phase model shows a large decrease in performance. 

These findings indicate similar results to the ones observed at window-level. Hence, using the Phase model we attained the largest percentage of correct trials in intrasession, whereas in intersession the Amplitude~+~Phase detector was the most reliable. In addition, the performance obtained using the Amplitude~+~Phase detector in intersession was similar to the one obtained in intrasession.

%The performance obtained as percentage of correctly classified trials were above chance level, which was computed by randomly interchanging the labels of a whole sequence of sliding windows (33 for the rest class and 8 for the pre-movement class) between classes. Due to class imbalance and to a long sequence length (41 windows), the chance level was 4\%.  
%Figure~\ref{fig:detection_models10}A shows the performance attained with the Amplitude model (left panel), Phase model (middle panel) and with the model based on the combination of Amplitude and Phase detection models output (right panel) for intrasession and intersession conditions in healthy subjects. Likewise, Figure~\ref{fig:detection_models10}B presents the results of the three detection models in stroke patients.  

%We conducted a statistical test using repeated measures two-way ANOVA, with factors ``detection model type'' (amplitude-, phase-  and ``evaluation type'', in which the average performance across sessions subjects' performance are the repeated measures. The goal of this statistical analysis was to measure the significance of the effects of the factors on performance. 

We present next the results of the repeated measures two-way ANOVA for the trial-based performance of the three detection models in the two evaluation conditions. 

In healthy subjects, Mauchly's test indicated that the assumption of sphericity was not violated neither for the model type factor, nor for the evaluation type factor.

All effects are reported as significant at $p$~$<$~0.05. There was a significant main effect of the factor ``type of the detection model'' on performance, $F$(2,~18)~=~20.21, $p$~$<$~0.001.
There was also a significant main effect of the factor ``type of evaluation'' on performance, $F$(1,~9)~=~93.20, $p$~$<$~0.001.
Furthermore, we found a significant interaction effect between the type of detection model and the type of evaluation used, $F$(2,~18)~=~42, $p$~$<$~0.001. This indicates that the type of evaluation had different effects on the attained performance, depending on which type of detection model was used.
Comparing intrasession with intersession performance within each model, the post-hoc Bonferroni corrected $t$-tests showed that the Amplitude model and the Phase model performed significantly worse in intersession than in intrasession ($p$~$<$~0.001 for both models), whereas the performance of the Amplitude~+~Phase model was not significantly different between the two conditions. 
Similarly, comparing different models in the intersession condition, we found that the Amplitude~+~Phase model was significantly better than both the Amplitude model ($p$~=~0.001) and the Phase model ($p$~$<$~0.001). Taken together, these results show the advantage of the Amplitude~+~Phase model in the transfer between sessions. 
%Pairwise signed rank Wilcoxon tests and Bonferroni-Holm correction for multiple comparisons revealed significant differences between intrasession and intersession evaluations for the Amplitude model ($p$~$<$~0.001) and for the Phase model ($p$~$<$0.001), but not for the Amplitude~+~Phase model. For the intersession evaluation, significant differences have been found between the Amplitude~+~Phase model and the Amplitude model ($p$~=~0.001) and between the Amplitude~+~Phase model and the Phase model ($p$~$<$~0.001).
%These findings suggest that the Amplitude~+~Phase model is robust to intersession variability.

\subsubsection{Intersubject evaluation}
For the intersubject transfer, we report in Figure~\ref{fig:detection_models10}C similar trial-based performance using the Amplitude~+~Phase model (62.5\%) relative to the intrasession evaluation. However, using the single-view models the attained performance was below both intrasession and intersession evaluations.
The robustness of the multi-view model to session- and subject-specific variability is due to a better discrimination between the movement intention and rest classes provided by the output combination of the single-view classifiers.
We found no statistically significant difference according to a pairwise Wilcoxon signed rank test between intra- and intersubject evaluations using the Amplitude~+~Phase model in this group of healthy subjects. 
We provide all the performance results for all the conditions in healthy subjects in the Supplementary Materials. 

\subsubsection{Detection time evaluation}
Next, we performed the analysis of the detection time in intrasession for the three detection models. Based on the neurophysiological results presented in section~\ref{subsec:NeuroResults}, we surmised that the phase-based model will detect earlier the intention of walking compared to the amplitude-based model. 

Figure~\ref{fig:detection_models10}D presents for the healthy subjects, the output of three detection models as a probability of belonging to the intention of walking class, as well as the subject-specific detection time for the three detection models. All the models show an increase in the probability of gait intention before the onset of motion. 

%We measured, as shown in the bottom panel of Figure~\ref{fig:Latencies}A , the latency (time point of the first correctly classified window as walking intention) with the three detection models. 
For all the subjects the multi-view model had the earliest detection time ($910$~ms before the movement onset). 
%The difference between the detection time of the multi-view model and the Phase model was statistically significant. 
Moreover, for the Phase model and for the multi-view model we observe longer time intervals between the detection of gait intention and the actual movement onset compared to the Amplitude model.    
In average we measured a difference of $240$~ms between the output of the multi-view model and the single-view Amplitude model, and a difference of $175$~ms between the Phase model and the Amplitude model. Furthermore, according to the paired signed rank Wilcoxon test and after Bonferroni-Holm correction for multiple comparisons, we found statistically significant differences between the latencies of the models ($p$~=~0.0078 between the Amplitude~+~Phase model and the Amplitude model, and $p$~=~0.0117 for the other two model comparisons). %)\footnote{~The order of the values on which the~\textit{p}-value is based was the same for all the subjects}.
%We performed the same analysis, presented in Figure~\ref{fig:Latencies}B, for the stroke patients, and found statistically significant difference between the latencies of the Phase model and Amplitude model, as well as between the multi-view model and the Amplitude model latencies ($p$~=~0.0236). %\textsuperscript{2}
%In average we found a difference of $251$~ms between the output of the multi-view model and the Amplitude model for the stroke patients group, and a difference of $210$~ms between Phase and Amplitude models latencies. 

\subsection{Neurophysiological analysis of stroke patients}
\begin{figure*}
\centering
\includegraphics[width=1\textwidth]{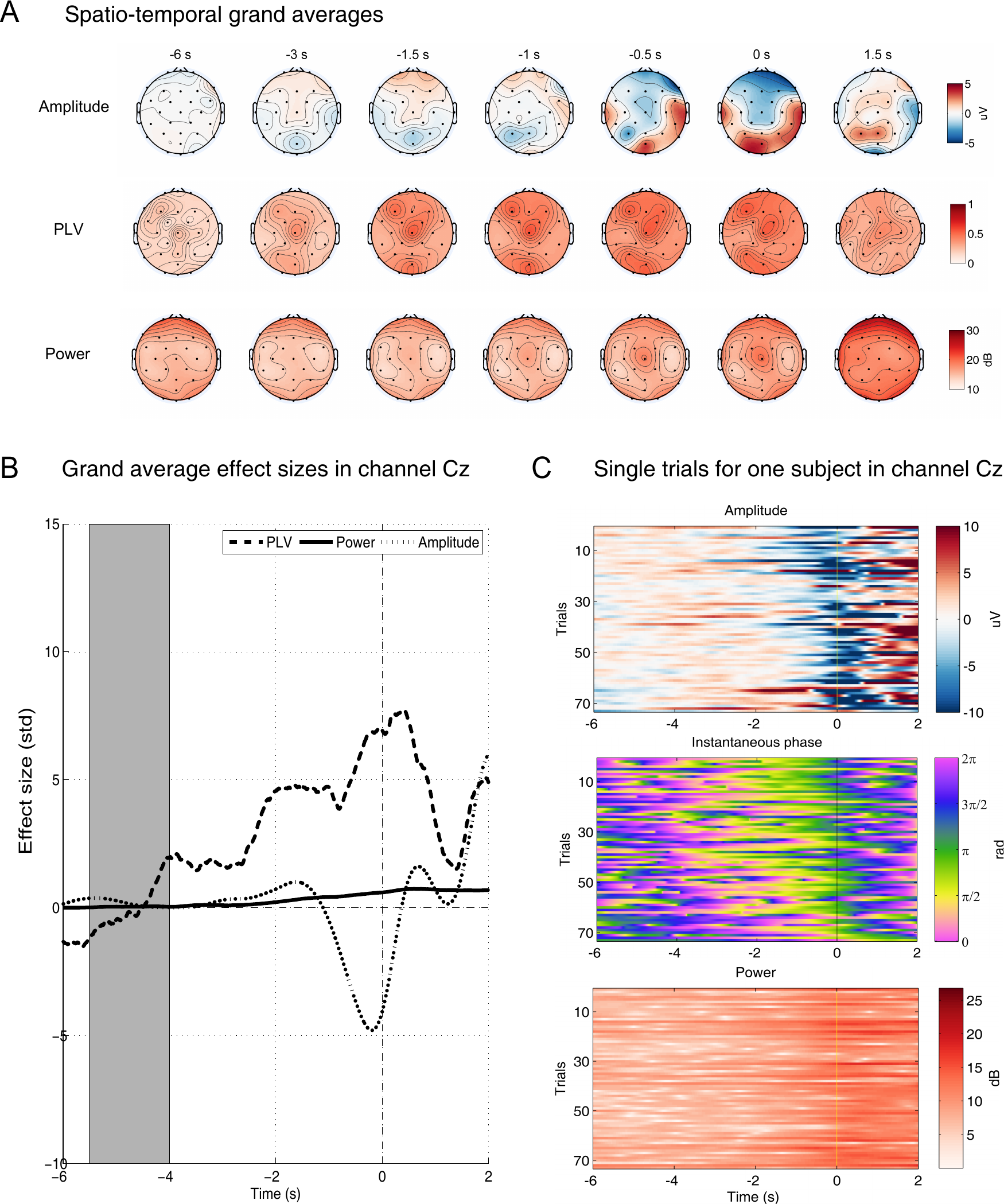}%{rev_stroke_neuro_corr.eps}%{Figure3_Neuro_strokes.eps} % Figure3_Neuro_strokes
\caption{Stroke patients neurophysiological results.
A. Grand average EEG correlates of gait intention in a time interval between $-6$ and $1.5$~s relative to the motion onset. 
Top panel: Movement related cortical potential (MRCP) amplitude representation; 
Middle panel: Phase-locking value (PLV) over task repetitions; 
Bottom panel: Power on decibels scale. 
B. Effect sizes for each of the three features quantified in standard deviation units in channel Cz for the grand average across patients. The gray area indicates the baseline used for the computation of the effect size.
C. Single trials illustration of the features in central channel Cz for one patient across multiple repetitions. 
The vertical dashed line indicates the onset of motion.}
\label{fig: Figure2_Neuro_Strokes}
\end{figure*}

%========= STROKES ====================
In Figure~\ref{fig: Figure2_Neuro_Strokes} we present the MRCP features in stroke patients (dataset 2). First, in Figure~\ref{fig: Figure2_Neuro_Strokes}A we observe these features in grand average over subjects. The MRCP amplitude shows a decrease in voltage, while the PLV presents an increase in magnitude over the motor cortex area with a large spread towards precentral and frontal areas. The power shows a slight increase in the proximity of movement onset. 

In Figure~\ref{fig: Figure2_Neuro_Strokes}B, we computed the grand average effect sizes of the three features relative to the baseline interval ($-5.5,-4$)~s. During the intention of motion interval ($-1.5,0$)~s, PLV has the largest average effect size (5.23 standard deviations), amplitude has a lower effect size (3.25 standard deviations) and power shows the smallest effect size (0.47 standard deviations). 
%
%In Figure~\ref{fig: Figure2_Neuro_Strokes}B, at the movement onset, PLV has the largest effect size, 7 standard deviations, while the amplitude has a lower effect size of 4.9 standard deviations. Power shows the smallest effect size of only 0.6 standard deviations. 
We evaluated the statistical significance between the absolute value of the averaged effect size of three representations (amplitude, instantaneous phase and power) in the intention of motion interval ($-1.5,0$)~s using a pairwise Wilcoxon test.
After the Bonferroni-Holm correction we found a statistically significant difference between instantaneous phase and power ($p$~=~0.004), as well as between amplitude and power ($p$~=~0.039). However, the difference between instantaneous phase and amplitude was not statistically significant ($p$~=~0.734).

The illustration of the behavior in single trials of the three features is depicted in Figure~\ref{fig: Figure2_Neuro_Strokes}C over the central channel Cz, for one representative patient.  
The top panel shows the MRCP amplitude features aligned relative to the movement onset. A negative deflection is found starting $-1.5$~s relative to the onset. 
We observe a large inter-trial variation in amplitude and in instantaneous phase. 
We found a slight increase in power starting $0.5$~s before the movement onset and lasting during the movement.

\subsection{Evaluation of detection models in stroke patients}
\subsubsection{Intra- and intersession evaluation}
As in the case of healthy subjects, we analyze first the performance in the intra- and inter-session conditions using the two metrics of performance.

\paragraph*{Window-based performance}

\begin{figure*} % Stroke
\begin{tabular}{cc} % % FIGURE 4 a b
  \includegraphics[scale = 1]{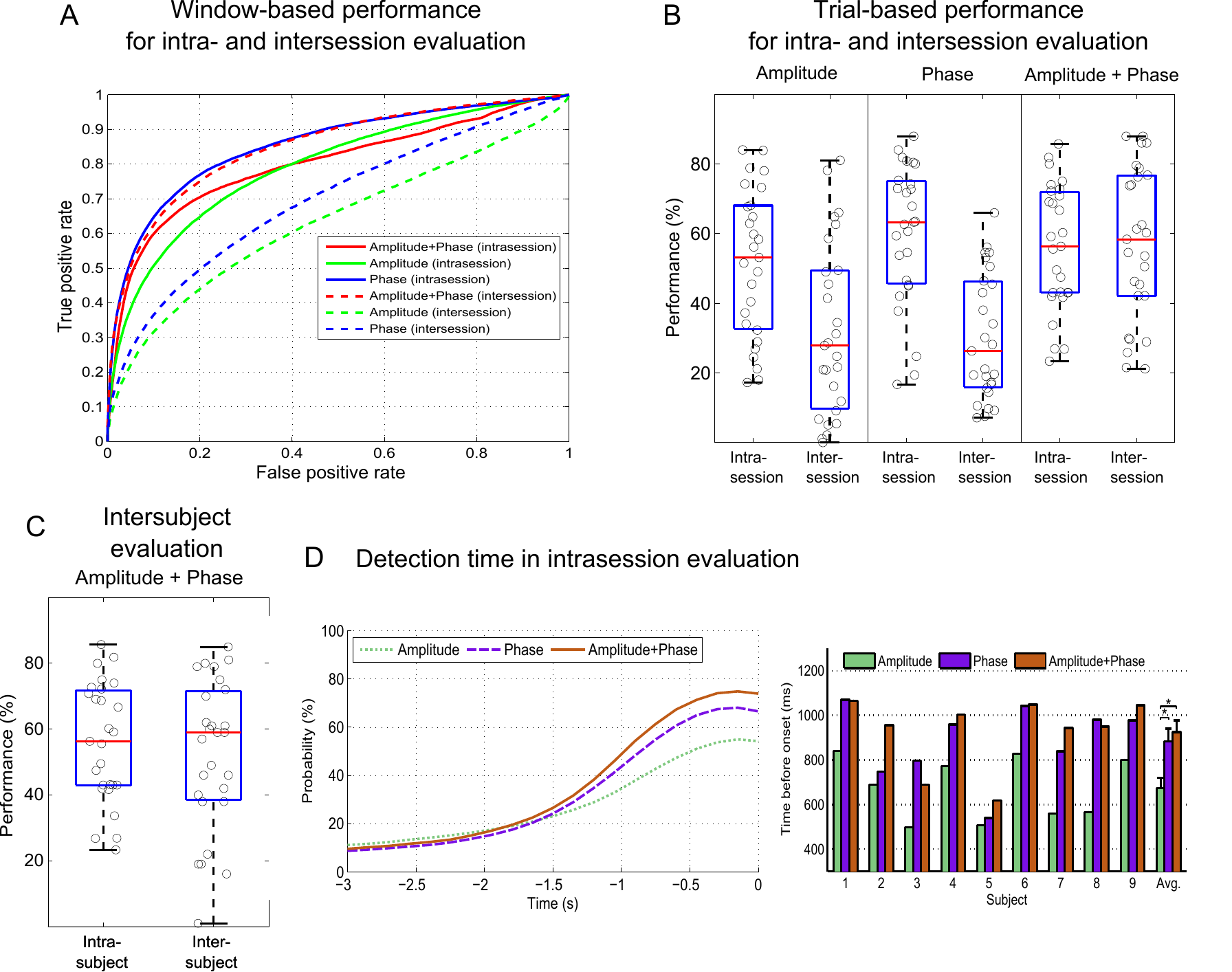}%{rev_Figure_det_stroke_1.eps}%{rev_Fig_perf_stroke.eps} %\qquad   \includegraphics[width=0.46\textwidth]{Figure4b_Decoding_strokes.eps} \\ 
\end{tabular}
\caption{Performance detection of gait intention in stroke patients using three models: Amplitude, Phase and Amplitude~+~Phase, in intrasession and intersession evaluations, computed in intra- and intersession with two metrics: 
A. window-based metric and B. trial-based metric. The circles mark subject and session specific performance. C. Trial-based intra- and intersubject evaluation. D. Detection time evaluation of the three detection models in intrasession condition. 
}
\label{fig:detection_models10_stroke}
\end{figure*}

In Figure~\ref{fig:detection_models10_stroke}A the performance is presented as averaged ROC curves across patients for the three detection models, in two evaluation conditions, intrasession and intersession. 
%To quantify the window-based performance in our unbalanced experimental design,
We report the area under the curve (AUC) as a summary statistic of the presented ROC curves.
During intrasession the highest averaged AUC was attained for the Phase model ($M$~=~0.85, $SD$~=~0.12), while the lowest AUC was obtained using the Amplitude model ($M$~=~0.79, $SD$~=~0.09).
For the intersession, the highest AUC was obtained using the multi-view model ($M$~=~0.84, $SD$~=~0.12), and the lowest AUC was attained using the Amplitude model ($M$~=~0.63, $SD$~=~0.12).
We attained the average empirical chance level of 58$\pm$2\%, with Cohen's $\kappa$~=~0.63$\pm$0.12.
%
%The obtained results in stroke patients are similar to the ones reported previously in healthy subjects.  Hence,
We conclude that in the group of stroke patients, the Phase model attains the highest performance in intrasession, while in intersession the Amplitude~+~Phase model is the most reliable for the window-based classification relative to the other detection models.  

\paragraph*{Trial-based performance}

Figure~\ref{fig:detection_models10_stroke}B shows the performance of the three detection models at trial level, as percentage of correctly classified trials, in intra- and intersession conditions.

In stroke patients, during intrasession condition, the Phase model attains the highest trial-based performance ($Mdn$~=~63.3\%, range~=~16.6\% - 88\%), followed by the Amplitude~+~Phase model ($Mdn$~=~56.3\%, range~=~23.3\% - 85.8\%) and by the Amplitude based model ($Mdn$~=~53.1\%, range~=~17.2\% - 84\%).

During the intersession transfer condition, the multi-view model attained similar performance ($Mdn$~=~58.3\%, range~=~21.1\% - 87.9\%) compared to the intrasession evaluation. 
The intersession transfer performance attained with the Amplitude model and with the Phase model shows a large decrease in performance.

%The median score attained using the Phase model for intersession transfer is 26.3\% in a range between 7.1\% and 66\%.
%The median score for the Amplitude model in intersession transfer is 27.8\% in a range between 0\% and 81\%. 
%The multi-view model yield a 56.3\% median performance in a range between 23.3\% and 85.8\% during intrasession. Furthermore, 
%During the intersession transfer condition, the multi-view model attained 58.3\% median performance in a range between 21.1\% and 87.9\%. 

We conducted the same statistical analysis using repeated measures two-way ANOVA, as previously described in healthy subjects.
For the stroke patients, Mauchly's test indicated that the assumption of sphericity was not violated for neither of the effects of the type of model (Amplitude, Phase or Amplitude~+~Phase) nor for the effects of the type of evaluation (intra- or intersession). 

%All effects are reported as significant at p < .05. 
There was a significant main effect of the factor ``type of the detection model'' on performance, $F$(2,~16)~=~15.57, $p$~$<$~0.001.
There was also a significant main effect of the factor ``type of evaluation'' on performance, $F$(1,~8)~=~35.17, $p$~$<$~0.001 and a significant interaction effect between the type of detection model and the type of evaluation used, $F$(2,~16)~=~9.89, $p$~$<$~0.001. This indicates that the type of evaluation had different effects on the attained performance, depending on which type of detection model was used.
Bonferroni-Holm corrected $t$-tests revealed significant differences between intrasession and intersession evaluations for the Amplitude model ($p$~=~0.010) and for the Phase model ($p$~=~0.034), but not for the Amplitude~+~Phase model. For the intersession evaluation, significant differences have been found between the Amplitude~+~Phase model and the Amplitude model ($p$~=~0.002) and differences close to significance between the Amplitude~+~Phase model and the Phase model ($p$~=~0.058). 
These findings suggest that the Amplitude~+~Phase model is robust to intersession variability and it attains the highest scores compared to the single-view models.

\subsubsection{Intersubject evaluation}

For the intersubject transfer we obtained similar trial-based performance using the multi-view model (59\%) relative to the intrasession evaluation, as shown in Figure~\ref{fig:detection_models10_stroke}C. The attained accuracy for single-view models was below the accuracy of both intrasession and intersession evaluations.
The multi-view model has a lower sensitivity to the specific threshold obtained in cross-validation and, consequently, is less sensitive to changes in the features distributions. 
We found no statistically significant difference according to a Wilcoxon signed rank test between intra- and intersubject evaluations using the Amplitude~+~Phase model.
We provide all the performance results for all the conditions in stroke patients in the Supplementary Materials. 

\subsubsection{Detection time evaluation}
In addition, we performed the detection time analysis of the three detection models in the intrasession evaluation.
% Based on the neurophysiological results presented in section~\ref{subsec:NeuroResults}, we surmised that the Phase based model will detect earlier the intention of walking compared to the Amplitude based model. 
Figure~\ref{fig:detection_models10_stroke}D presents for the stroke patients, the output of the detection models as a probability of belonging to the intention of walking class, as well as the subject-specific detection time for the three detection models. All the models show an increase in the probability of gait intention before the onset of motion. 
For all the subjects the multi-view model had the earliest detection time ($924$~ms before the movement onset). 
We found statistically significant differences according to a signed rank Wilcoxon test and after Bonferroni-Holm correction, between the detection time of the Phase model and Amplitude model, as well as between the multi-view model and the Amplitude model detection time ($p$~=~0.0236). %\textsuperscript{2}
In average we found a difference of $251$~ms between the output of the multi-view model and the Amplitude model for the stroke patients group, and a difference of $210$~ms between the detection time of the Phase and Amplitude models.

%---------------------
\section{Discussion}
%======= Summary of the study=======
This paper studies the instantaneous phase representation of the MRCP and its usage for the detection of gait intention. 
%This study presents the phase representation of MRCP amplitude correlate of gait intention, and focuses on the contribution of MRCP instantaneous phase features to successful detection of gait intention. 
%
We show that the MRCP instantaneous phase has a higher signal-to-noise ratio and that its pattern related to movement intention emerges earlier than the patterns of the MRCP amplitude or power. 
Next, we propose three detection models: two single-view models, one based on MRCP amplitude features and another based on MRCP instantaneous phase features, and a 
%a single-view model based on MRCP amplitude features, a single-view model based on MRCP instantaneous phase features,  
multi-view model that combines the outputs of the single-view models. 
%MRCP amplitude and MRCP instantaneous phase models. 
During intrasession evaluation the single-view Phase model attained the highest detection score compared to the other detectors. However, in both intersession and intersubject transfer conditions,  the multi-view model outperformed the single-view models.   

\subsection{Neurophysiological results}
In recent studies~\cite{ng2013eeg,klimesch2004phase,gruber2005alpha} the instantaneous phase of theta and alpha frequency bands has shown a higher signal-to-noise ratio and earlier patterns compared to the ones observed in power.  
Furthermore, Ng et al.~\cite{ng2013eeg} have shown that these phase patterns persist even in the absence of an increase in power.

%%Phase dynamics have been shown to contain information about the process of interest. 
%Phase patterns have been shown to anticipate the appearance of expected auditory and visual stimuli. 
%%====Phase has more information than power and it has a specific pattern in the absence of an increase in power=====
%In a recent study, Ng et al.~\cite{ng2013eeg} showed that the phase pattern of theta oscillations contains more information than power. Furthermore, they have shown that this pattern persists in the absence of an increase in the power of the oscillation.  
In our analysis we show that the phase patterns of the slow oscillations in delta frequency band related to the movement intention have the highest signal-to-noise ratio and appear even earlier than $1.5$~s before the movement onset, which is conventionally used as a lower bound for the effect of movement intention to appear in EEG~\cite{deecke1976voluntary} (see Figure~\ref{fig: Figure1_Neuro_Healthy}B and Figure~\ref{fig: Figure2_Neuro_Strokes}B). 
%Moreover, they found that the information that can be discriminated by firing rates can also be discriminated by phase patterns, but not by power. Therefore, the auditory stimulus-specific phase pattern persists in the absence of an increase of oscillation power.
%In Figure~\ref{fig: Figure1_Neuro_Healthy}B and Figure~\ref{fig: Figure2_Neuro_Strokes}B we show that the PLV has the highest signal-to-noise ratio and an effect of the movement intention is apparent the earliest in PLV. Moreover, this effect is evident even earlier than $1.5$~s before movement, which is conventionally used as a lower bound for the effect of movement intention to appear in EEG~\cite{deecke1976voluntary}.
%that it appears in the absence of an increase in power, before the movement intention (a priori defined to start around $1.5~s$ before the movement onset~\cite{deecke1976voluntary}).  
Next, we show that the increase in the power of the slow (delta) oscillations during the intention of movement interval is relatively small and it appears after the phase pattern (see Figure~\ref{fig: Figure1_Neuro_Healthy}C  and  in Figure~\ref{fig: Figure2_Neuro_Strokes}C). 

%====Phase has an earlier pattern than amplitude====
%
As previously stated in~\cite{klimesch2004phase, gruber2005alpha} theta and alpha oscillations reset early relative to the stimulus presentation (already before or around $50$~ms). 
More precisely, a phase reset has been found at a time point when there is still no activity visible in power. %component of the event-related potential visible in amplitude. 
%More precisely, they studied the generation of P1-N1 complex due to alpha and theta phase locking oscillations. 
Recent studies~\cite{stefanics2010phase, saleh2010fast} 
%present the anticipatory behavior of delta phase locking oscillations to attended rhythmic auditory or visual stimuli. They 
have shown that phase synchronization of slow (delta) oscillation depends on the likelihood of the appearance of a significant event, suggesting that phase locking is not only a mechanistic consequence of periodic stimulation. 
Based on previous findings~\cite{stefanics2010phase, saleh2010fast} and on the self-paced nature of our protocol, we consider the significant event to be the subjects' intention to move whenever they want. We found a non-uniform pattern in the phase of the delta oscillation, related to the intention to move, preceding the movement onset.     
The phase synchronization value (PLV) anticipates the pattern elicited by the MRCP amplitude with more than half a second, both in healthy subjects (see Figure~\ref{fig: Figure1_Neuro_Healthy}A) and stroke patients (see Figure~\ref{fig: Figure1_Neuro_Healthy}B).

%====ERP models==================
In the last decade, a debate about the generation of event-related potentials (ERPs) has proposed two models. The evoked model~\cite{arieli1996dynamics} states that ERPs are generated by additive evoked responses which are completely independent of ongoing background electroencephalography (EEG). 
On the other hand, the phase reset model~\cite{makeig2002dynamic, tass2007phase, sauseng2007event} suggests a resetting of ongoing brain oscillations to be the neural generator of ERPs.

%====Phase resetting theory====
The presented neurophysiological results suggest phase resetting as a potential neural generator of MRCPs after the appearance of gait intention. 
In particular, phase resetting indicates that after each stimulus, the phase of a certain rhythm is shifted towards a dominant value in relation to the stimulus~\cite{tass2007phase}. 
More specifically, looking at the distribution of instantaneous phase over many trials in Figure~\ref{fig: Figure1_Neuro_Healthy}C for healthy subjects and in Figure~\ref{fig: Figure2_Neuro_Strokes}C for stroke patients, one can observe a distribution of random fluctuation in the instantaneous phase of the slow oscillations during the rest period (before the movement intention), changing after two seconds before the movement onset to a distribution that peaks about a dominant value from $\pi/2$ to $\pi$.

%====================================
It is important to note that the group of healthy subjects is not a control group for the stroke patients, since the two groups have been recorded in different datasets and there are several factors that differentiate them, such as age, medication, slight differences in the experimental protocol, etc.
Although a direct comparison between the two groups of subjects is not straightforward due to the aforementioned  factors, we observed similar patterns between the features of the two groups of subjects.  
The MRCP amplitude and PLV have a similar behavior in healthy and stroke subjects, although, a larger spread is observed in stroke patients covering frontal, precentral and central areas (see Figure~\ref{fig: Figure1_Neuro_Healthy}A and Figure~\ref{fig: Figure2_Neuro_Strokes}A).   
The effect sizes of the three representations have smaller magnitude in stroke patients relative to the ones attained in healthy subjects (see Figure~\ref{fig: Figure1_Neuro_Healthy}B and Figure~\ref{fig: Figure2_Neuro_Strokes}B).
Note that the behavior of the MRCP amplitude's effect size is conserved between the two groups of subjects, whereas the effect size of PLV and power features show dissimilarities between groups. 
The PLV in stroke patients presents the highest effect size in the proximity of the onset as observed in healthy subjects, but the increase is attenuated and more gradual in stroke patients than in healthy subjects. Power presents a different effect-size pattern between the two groups of subjects. Although increasing before the onset of motion for both groups, in stroke patients the effect-size of power remains constant also during movement.

%We observe in Figure~\ref{fig: Figure2_Neuro_Strokes}C an inter-trial variation in both amplitude (top panel) and instantaneous phase (middle panel). Although Figure~\ref{fig: Figure1_Neuro_Healthy}C and Figure~\ref{fig: Figure2_Neuro_Strokes}C present individual subjects, it is instructive to generalize these findings since the reported subjects are representative for each of the groups. Therefore, the inter-trial variation could be partially explained by the differences in onset selection between the two groups, leading to a higher level of noise (tremor, involuntary muscle contractions, etc.) in the EMG signal recorded in patients compared to healthy subjects, as well as  due to several factors such age or different brain plasticity changes. 
%Moreover, in the bottom panel, the power representation has a large attenuation compared to the representation of power in the healthy subjects group.  

%Further research is needed to establish the effect of lesion on the neural correlates of movement, by comparing a group of stroke patients with their healthy control group. 

In the future, a comparison should be performed between a group of stroke patients and their matched healthy control group in order to establish the effect of lesion on the dynamics of the neural correlates of movement intention.  
In addition, MRCP phase patterns could be analyzed under different protocols that require the execution of different tasks (such as tongue, hand, foot movements, etc.). We consider that several characteristics of the MRCP investigated in amplitude (such as latelarization, preference for specific areas of motor cortex, etc.) could be further explored by the dynamics of the MRCP instantaneous phase.

\subsection{Detection models evaluation}
MRCPs have been investigated in many studies for the detection of upper limb, lower limb and gait movement intention~\cite{niazi2011detection, lew2012detection, do2006movement, niazi2013detection, garipelli2013single, jiang2014brain, sburlea2015continuous, sburlea2015detecting}. All these studies analyzed MRCP amplitude representation, whereas MRCP phase oscillatory activity, in delta band has not yet been evaluated for the detection of gait intention.  

However, the phase synchronization of oscillatory signals in mu band has been investigated to assess connectivity between brain areas while performing imaginary movements~\cite{daly2012brain, wei2007amplitude, he2012combination, hamner2011phase, hsu2014enhancing}. They concluded that phase patterns contain complementary information to the power in the mu band and can effectively improve classification. 
%Using linear discriminant analysis,  Hsu~\cite{hsu2014enhancing} attained an increase from 75.6\% to 80.7\% in controlling a bar with imaginary hand movements, by including phase features. 
%Hamner et al.~\cite{hamner2011phase} showed that the metrics of synchronization derived from instantaneous phase features have a comparable performance relative to the power of mu band.
%In another study~\cite{he2012combination}, He et al. combined the amplitude information encoded by common-spatial pattern (CSP) features of mu band with phase synchronization information given by PLV features. They attained an average increment of 5.4\% to the results obtained using only amplitude information as CSP features, for the classification of different imaginary hand movements. 
%Moreover, Wei et al.~\cite{wei2007amplitude} show that PLV features and autoregressive coefficients from mu band contain complementary information and can effectively improve classification of different types of imaginary movements. 
%
Our findings show that using either of the metrics (window-based or trial-based) the model built on MRCP phase features attained the largest performance in the intrasession condition in both healthy subjects %($M$~=~0.88, $SD$~=~0.09) 
and stroke patients. % ($M$~=~0.85, $SD$~=~0.12). 
%Similarly, using the trial-based metric the model based on MRCP phase features attained in intrasession condition the highest median number of correctly detected intentions of walking in healthy subjects ($Mdn$~=~66.5\%, range~=~11\% - 87\%) and in stroke patients ($Mdn$~=~63.3\%, range~=~16.6\% - 88\%). 
Using the trial-based metric we observed an improvement of 4.5\% in healthy subjects and of 10.1\% in stroke patients compared to the model based on more conventional MRCP amplitude features. 
%The analysis of stroke patients yields a median improvement of 10.1\% when using the MRCP phase model compared to the MRCP amplitude model.  
In a recent study~\cite{jiang2014brain} that investigates the amplitude of the MRCP for the detection of gait intention in healthy subjects, authors report a TPR of 76.9\% and a FPR of 2.93~$\pm$~1.09 per minute. We can compared these results with our intrasession window-based performance, in which for a similar FPR we attained a TPR of 87.6\% using the Phase model, followed by the Amplitude~+~Phase model (TPR~=~81.9\%) and by the Amplitude model (TPR~=~79.2\%). The reported TPRs are derived from the AUC reported in Figure~\ref{fig:detection_models10}A and Figure~\ref{fig:detection_models10_stroke}A.

% === SVM as a more appropiate classifier for nonlinear data, such as the circular distrbution of phase features.
The majority of EEG based BCIs that study MRCP information for the detection or classification of movement use either linear discriminant analysis or template matching methods ~\cite{niazi2011detection, lew2012detection, do2006movement,jiang2014brain, sburlea2015continuous}. 
We used SVM classification models for the detectors based on both MRCP amplitude and MRCP instantaneous phase.
The benefit of the SVM classifier was shown in the classification of instantaneous phase features, which have a nonlinear representation. Therefore, a nonlinear classifier such as RBF-SVM can be more successful in separating data that is not linearly discriminable.  
However, since amplitude features are linearly separable, the results attained with the MRCP amplitude detector were similar to the ones obtained with sparse linear discriminant analysis, shown in~\cite{sburlea2015continuous}.

%
%%====Latencies of the proposed models.
In BCI based neurorehabilitation, plasticity can be induced if the response latency of the detector relative to the user's movement intention is in the order of a few hundred milliseconds or smaller~\cite{xu2014enhanced}.
In the current study we show, for the two groups of subjects, that the detection models based on phase features  attained an earlier detection compared to the model based on amplitude features.  
It would be interesting to evaluate the effect of such an earlier detection time, that precedes the movement onset, on plasticity and functional recovery.
%The benefits of an earlier detection are two-fold. First, 
From a technological perspective, an earlier detection yields a longer time interval from the detection point until the decision-making process to trigger the rehabilitation device, allowing the integration of other processes. 
%Furthermore, it would be interesting to evaluate the effect of an even earlier detection time of the movement intention on plasticity and functional recovery.
 
Previous studies that use MRCP amplitude information for the detection of movement intention label the interval from $1.5$~s prior to the movement onset as the intention of motion interval or the pre-movement state~\cite{deecke1976voluntary, lew2012detection, jiang2014brain, bai2011prediction, mrachacz2014novel, sburlea2015continuous}. In the current study, we used the same convention. 
However, we observed that the MRCP instantaneous phase pattern emerges earlier than the MRCP amplitude or power representations. 
Therefore, we surmise that the earlier appearance of the phase pattern might increase the accuracy of the detector by using different relaxation and pre-movement state labeling criteria in future studies.

%====TRANSFER.
A common trend in BCI to reduce (or even eliminate) the time needed for recalibration and increase robustness to artifacts and nonstationarities is the integration of data from multiple sessions and subjects (see~\cite{tu2012subject, lotte2010learning, krauledat2008towards,arvaneh2012omitting, arvaneh2013eeg,fazli2009, fazli2009a,fazli2011l, fazli2015learning} for motor imagery,
~\cite{garipelli2013single, sburlea2015continuous, sburlea2015detecting, sburlea2015adaptation, lopez2015brain} for real movement execution 
and~\cite{lottesignal} for workload and mental imagery). 
The calibration of the BCI is usually associated with significant costs and subjects tiredness~\cite{belda2011rehabilitation}. 
The current study shows, for the detection of gait intention, that the combination of MRCP amplitude and phase information achieves similar performance during intrasession to the performance attained without recalibration during intersession and intersubject transfer conditions ($\sim$61\% in healthy subjects and $\sim$57\% in stroke patients). 
In a previous study, we reported similar results between sessions for healthy subjects, but with lower overall performance ($\sim$58\%)~\cite{sburlea2015continuous}. Furthermore, in stroke patients it was still necessary to integrate small amounts of data ($\sim$10~minutes) for every session to avoid large decay in performance~\cite{sburlea2015adaptation}.  
Despite that the results of the detector based on phase and amplitude features suggest that session- and subject-specific BCI recalibration could be unnecessary for the detection of gait intention, there is still a need to understand where is the limit of the proposed combination of features. First, longer term usage of the BCI may impose larger changes in the feature distributions that can affect performance (e.g. due to neural plastic changes during rehabilitation and to the learning process). Second, results may vary for other types of movement (e.g. motor imagery patterns are known to have smaller amplitude than the ones for real movements). Moreover, in~\cite{morash2008classifying} they have shown that upper limb, tongue and lower limb movements have different activity patterns.
For these cases, the proposed models could be useful to allow long term operation in a wide range of movements.

%=====
%For future work: 
Traditional BCIs use a range of features to allow the user to control an external device, such a computer or a prosthesis~\cite{wolpaw2002brain}. 
These feature types can include amplitude values, frequency based features such as band powers and time-frequency maps of cortical activity of specific brain areas~\cite{lotte2007review}. 
Our results suggest that phase is a better representation of MRCP amplitude that improves the detection of movement intention and removes the need of session- and subject-specific BCI recalibration widening the integration of this type of detectors in other applications that need an earlier assessment of movement intention for the control of an external device. 
Due to the robustness in transfer learning we consider gait rehabilitation as a primary application of this technique, however other applications in which the users perform various types of movement (e.g. upper limb, lower limb movements) could also benefit from off-the-shelf BCI technology.

\section*{Acknowledgments}
Authors thank Filip Melin\v{s}\v{c}ak, Neboj\v{s}a Bo\v{z}ani\'{c}, Mario Mulansky and Thomas Kreuz for valuable discussions and to Isabel Alquacil Diego and Roberto Cano de la Cuerda for their help during the  recordings with stroke patients, as well as for the assessment of patients' condition.
% and careful reading of the paper and 
Andreea Ioana Sburlea would like to thank the Institute of Complex Systems, CNR in Sesto Fiorentino for hospitality during her research visit.
Authors acknowledge funding by the European Commission through the FP7 Marie Curie Initial Training Network 289146, NETT: Neural Engineering Transformative Technologies and the Spanish Ministry of Science projects HYPER-CSD2009-00067 and DPI2011-25892.

%\balance
%\addtolength{\textheight}{-12cm}   % This command serves to balance the column lengths
%%                                  % on the last page of the document manually. It shortens
%%                                  % the textheight of the last page by a suitable amount.
%%                                  % This command does not take effect until the next page
%%                                  % so it should come on the page before the last. Make
%%                                  % sure that you do not shorten the textheight too much.

\bibliographystyle{IEEE}
\bibliography{Ref_Phase}

\end{document}